\documentstyle[11pt]{article}
\input epsf
\def\ll{\label}
\def\re{\ref}
\def\c{\cite}

\def\r1{(\ref{$1})}
\def\ot{\otimes}

\def\th{\theta}
\def\ba{\begin{array}{c}}

\def\ea{\end{array}}

\def\da{\dagger}
\def\De{\Delta}

\def\ov{\over}
\def\ha{{1\over 2}}

\def\l{\left}
\def\l({\left(}
\def\r){\right)}
\def\r{\right}
\def\rw{\rightarrow}

\def\la{\lambda}
\def\al{\alpha}

\def\be{\begin{equation}}
\def\bc{\begin{center}}
\def\ec{\end{center}}
\def\bit{\begin{itemize}}
\def\eit{\end{itemize}}
\def\ee{\end{equation}}
\def\ed{\end{document}}
\def\bea{\begin{eqnarray}}
\def\eea{\end{eqnarray}}
\def\efr{\end{flushright}}

\begin{document}
\title{
 Yang-Baxter algebra  and generation of  quantum integrable models
}
%\vskip 1cm

\author{
Anjan Kundu \footnote {email: anjan.kundu@saha.ac.in} \\
  Saha Institute of Nuclear Physics,\\
 Theory Group \& Centre for Applied Mathematics and Computer Science, \\
 1/AF Bidhan Nagar, Calcutta 700 064, India.
 }
\maketitle
\vskip 1 cm

%\vskip 4 cm
%\e{itemize}
\begin{abstract}
%---------------------------------------------------
An  operator deformed quantum algebra
is discovered exploiting the
quantum  Yang-Baxter
equation with trigonometric R-matrix.
This novel Hopf algebra along with its $q \to 1$ limit
appear to be the most general Yang-Baxter algebra underlying
quantum integrable systems.
Three different directions of  application of this algebra in integrable systems
depending on different  sets of values of  deforming operators
are identified. Fixed  values on the whole lattice
yield subalgebras linked to standard quantum
integrable models, while the associated Lax operators generate and classify them
 in an unified way. Variable values
construct a new series of  quantum integrable inhomogeneous
models. Fixed but different values
at different lattice sites can produce a
novel class of  integrable   hybrid models including integrable
matter-radiation models
and quantum field models with defects, in particular, a new quantum
integrable sine-Gordon model
with defect.
\end{abstract}

 PACS numbers 02.30.Ik,
%integrable systems
  02.20.Uw,
%quantum groups
05.50+q,
%lattice theories +statistics
03.65.Fd
% algebraic methods


\section{Introduction }
 Quantum integrable systems (QIS) are nonlinear, interacting and nonperturbative
quantum many body or field  models having large
symmetries with
mutually commuting set of conserved operators $\{C_n\}, \ n=1,2,\ldots$.
%$[C_a,C_b]=0, a,b=1,2,\ldots$.
 Such models usually  allow
 exact solution of  the
 eigenvalue problem through the algebraic
 Bethe ansatz method, for
   all  conserved operators  including the Hamiltonian of the system.

%\medskip
Every lattice  QIS is
represented by its own  Lax operator $L_j(\la)$
at each lattice site $ j=1,2, ...,N $ and for
the integrability  it should  satisfy quantum Yang-Baxter equation (QYBE)
\be  R(\la-\mu)L_j(\la)\otimes L_j(\mu)=
(L_j(\mu)\otimes I) (L_j(\la)\otimes I)R(\la-\mu), \ \ j=1,2, ...,N \ll{qybe}\ee
together with the ultralocality condition
\ $  L_j(\la)\otimes L_k(\mu)=
(L_k(\mu)\otimes I) (L_j(\la)\otimes I), \ \ $ at $ \ k \not =j$.
These two relations combined together lead to several important consequences.

i) They provide sufficient condition for quantum integrability.
 Defining monodromy matrix $T(\la)= \prod_j
L_j(\la) $  on the whole lattice,
 one  shows that the same QYBE holds also for this global object
:
\be  R(\la-\mu)T(\la)\otimes T(\mu)=
(T(\mu)\otimes I) (T(\la)\otimes I)R(\la-\mu).\ll{gqybe}\ee
Taking trace from both sides one derives the trace identity:
 $[\tau(\la),\tau(\mu)]=0,\ $ for $ \ \tau(\la)=Tr (T (\la)) $, yielding the
crucial  integrability condition $[C_n,C_m]=0 $, through
% conserved operators obtained from the expansion 
$\tau(\la)=\sum_n C_n \la ^n $

ii) QYBE (\ref{gqybe}) plays important role in exact solution through the
algebraic Bethe ansatz
method  by  providing commutation relations between
diagonal (generator of conserved operators) and off-diagonal ('creation' /
'annihilation' operators )
 elements of $T(\la)$-matrix.

iii) Finally,  QYBE (\ref{qybe})  defines the
commutation relations between  operator-elements of the quantum Lax
operator  ({Yang-Baxter (YB) algebra}),
specified by the quantum $R$-matrix,   a c-number matrix which fixes the
the structure constants of the algebra. Interestingly,   the very fact
that the same QYBE is valid  also at the global level (\ref{gqybe}), tells
that
this YB algebra
must be  a  Hopf algebra  
   $A $  with  defining properties:

{\it  coproduct}: \ $ \De:
A \to A \otimes A$
, \ \ {\it product}: m:\ $A \otimes A  \to A$,
\ \
{\it unit}:
 $e:
 k \to A
$,
 \ \
{\it counit}
: $\epsilon :
 A \to k
$
\ \
and {\it antipode} (inverse):  S:
$ A \to A$.

For some integrable systems when the last property is absent
 the algebra  becomes simply a bi-algebra (algebra +coalgebra).
The underlying YB algebra
classifies the    quantum integrable
 models into three major classes : {\it rational,
trigonometric} and {\it elliptic}, distinguished by
such  solutions  of the associated quantum $R $-matrix.
Here we will focus on the first two classes.

\section{Known Yang-Baxter algebra}
 The simplest  YB algebra is given by the standard
  Lie algebra
$ sl(2)$ :$ \  [ S^+ , S^- ]=  2 S^3 ,\ \ \ \  ~
 [S^3, S^\pm]  = \pm S^\pm .$
% \ll{sl2} \ee
 This is a  cocommutative Hopf algebra
generated by the  quantum  $R$-matrix:
\begin {equation}
R^{11}_{11} = R^{22}_{22}= a(\la),
\  R^{12}_{12} = R^{21}_{21}= b(\la), \ R^{12}_{21} = R^{21}_{12}= c
      ,    \ll{R-mat}\end {equation}
 having solutions $R^{rat}$ in rational functions of the
   spectral parameter:
$ \  a(\la)=\la+\al, \ \ b(\la)= \la , \ \
 c ={  \al}. \ $
% \ll{rrm}\ee
This algebra is linked to the well known quantum integrable models \c{kulskly}
like
 1) Nonlinear Schr\" odinger  (NLS) equation and the  lattice NLS, \
2) xxx-spin-$\ha$ chain\ \,
3) rational Gaudin model  etc.

A more  general  (and more interesting) YB algebra is
 associated with the trigonometric $R^{trig}$-matrix solution
of (\ref{R-mat}) as
$ \  a(\la)=\sin (\la+\al), \ \ b(\la)=\sin  \la , \ \
 c =\sin   \al. \ $
% \ll{trm}\ee
and is given by the celebrated quantum algebra
 $sl_q(2)$ $ \ 
%\be:
  [ S^+_q , S^-_q ]
=    \frac {\sin(\al 2S^3)}{\sin \al } ,\ \ \ \
 [S^3, S^\pm_q]  = \pm S^\pm _q, \ $
%\ll{slq2}\ee
where $q=e^{i \al} $ is the deformation parameter. FRT construction
\c{frt,fadrev}
  gives an elegant way of obtaining this quantum algebra
 together with  its non-cocommuting Hopf algebra properties and  the
explicit form of the associated $(2 \times 2) $ Lax operator.
$sl_q(2)$ was shown  \c{fadrev} to be the underlying YB algebra of
well known  quantum integrable systems like
1) sine-Gordon (SG) and lattice SG model, \
2) xxz-spin chain etc.

We however observe   that
 there exists a series of other important quantum integrable
models  of diverse nature having completely   different  Lax operators
but   associated with the same quantum matrix:
$R^{trig}$ or $R^{rat}$.  Few such
examples are the quantum Toda chain belonging to the rational class, and
the relativistic
Toda chain, Liouville model, derivative NLS etc.
  belonging to the trigonometric class.
Therefore in spite of the far-reaching success of the quantum algebra
 $sl_q(2)$ it is
 pertinent to ask

i)  Does it  exhaust all possible
 YB algebras associated with
the trigonometric  $R$-matrix ?
\
ii) What are the YB algebras underlying
 other integrable  models ?
\
iii) Can they all be obtained  from $sl_q(2)$ or $sl(2)$ ? 
\
iv) Can  there be any deep relation among
  diverse integrable  models sharing the same quantum $R$-matrix,
 despite
of having completely different  Lax operators ?

In search for  answers to these fundamental questions, we conjecture that
there must exist a more general YB algebra and
 the related    quantum Lax
operator  associated with the same
$R^{trig}$-matrix (and its $q\to 1 $ limit linked with
$R^{rat}$-matrix), which should  cover all quantum integrable models
with $(2 \times 2) $ Lax matrix. It should include
the known quantum algebras (and
 its  $q\to 1 $ limit)  along with  related models as  particular cases,
 while    having the freedom  to generate
  YB algebras underlying
 all other  known and new quantum integrable models.

\section { Operator-deformed  quantum
 algebra }

Supporting our conjecture we discover a conceptually novel YB algebra,
an algebra deformed  not only by usual parameter $q$, but also 
  by some operators.
   Our idea  is to follow  closely the
 FRT construction  for  
$sl_q(2)$ and the associated   Lax operator \c{frt},
but deform  the related Borel subalgebras further  by
 operators $\hat c^\pm_a, a=1,2 $, resulting   a new
 Lax operator
\be
L_{anc}^{trig}(\xi) = \left( \begin{array}{c}
  \xi{\hat c_1^+} e^{i \al s^3}+ \xi^{-1}{\hat c_1^-}  e^{-i \al s^3}\qquad \ \
2 \sin \al  s_q^-   \\
    \quad
2 \sin \al  s_q^+    \qquad \ \  \xi{\hat c_2^+}e^{-i \al s^3}+
\xi^{-1}{\hat c_2^-}e^{i \al s^3}
          \end{array}   \right), \quad
          \xi=e^{i \la}, \ll{aL} \ee
% $L^\pm(s_q^\mp,c^\pm_a,s^3)$,
 while  the $R^{trig}(\la)$-matrix remains the  same.
Inserting  (\ref{aL}) together with
  in QYBE
(\ref{qybe}) we derive the novel
  {\it operator}-deformed quantum algebra, we  call
{\it Ancestor algebra} \c{aa}, defined through the relations
\be [ s_q^ {+}, s_q^{-} ] =
 \left (\hat  M^+ \sin (2 \al s^3) -i
\hat  M^-  \cos
( 2 \al s^3  ) \right){1 \over \sin \al}, \quad
[s^3,s_q^{\pm}] = \pm s_q^{\pm} , \ \ [ \hat  M^\pm, \cdot]=0
 \ll{ancAlg} \ee
where the deforming operators
$\hat  M^\pm=  \frac {  1} {2 }  
(\hat c^+_1\hat c^-_2 \pm \hat  c^-_1\hat c^+_2 )$
 and all operators $\hat c^-_a \pm , \ a=1,2$ are mutually commuting as well
as  central
 (commuting with all other generators of algebra (\ref{ancAlg})).
We believe that  (\ref{aL}) is the most general form of the
 Lax operator, which can generate  $2\times 2 $
Lax operators of all integrable systems belonging to this class and
 the  operator-deformed    quantum
 algebra (\ref{ancAlg}) is the most general YB algebra allowed by
 the simplest   $R^{trig}$-matrix. This ancestor algebra
exhibits  the following
 unique and distinguishing  properties.

1). The  operator deformed quantum algebra (\ref{ancAlg})
  is a bialgebra (algebra +
coalgebra) with the
{\it Hopf algebra  properties}:

 i) \ {\it Coproduct} $\Delta$ :
 \[  \De (s_q^+)= s_q^{+} \otimes
\hat c^-_{1}q^{-s^3 }  +
    \hat  c^-_{2} q^{s^3 }\otimes {s_q^+}, \ \ \    \De (s_q^-)= {s_q^-} \otimes
\hat c^+_{2}q^{-s^3 }   +
   \hat  c^+_{1}q^{s^3 }\otimes {s_q^-}, 
\] \[
\Delta( { s^3})= I \otimes { s^3}+s^3 \ot I,
 \ \ \ \quad
\Delta(\hat   c_a^{\pm })=\hat c_a^{\pm } \otimes \hat  c_a^{\pm }, \ a=1,2 \]
Note that  unlike  deforming
 parameter q,  deforming operators $\hat  c_a^{\pm } $ and hence $\hat M^\pm $
 have nontrivial coproduct.

 ii)\ {\it Antipode} S:
\[ 
  S (  s_q^-)= - (\hat c_1^{+ })^{-1} e^{- i \al s^3 }s_q^- e^{i \al s^3
 }(\hat c_2^{+ }
)^{-1} , \ \
  S ( s_q^+ )=
 - (\hat c_2^{- })^{-1}e^{-i \al s^3 } S^+ e^{i \al s^3 }(\hat c_1^{- })^{-1},
\] \[
S (\hat c_a^{\pm })  =
(\hat c_a^{\pm })^{-1} , \ \ \ \qquad  S (e^{\pm i \al s^3 })  =
e^{\mp i\al s^3 }. \]
  %(7)
 In a QIS if any  $\hat c^\pm_a$=0, the antipode disappears
turning the    Hopf algebra  into a bialgebra.

iii)  {\it Counit} $ \epsilon $:  
\[ \epsilon (s_q^\pm) = 0, \    
     \ \epsilon (e^{\pm i \al s^3 })=1, \ \epsilon  (\hat c_i^{\pm }) = 
c_i^{\pm }  \]
For {\it product}  ${\sl m}$
 one can take the formal definition of multiplication in the algebra, while
 {\it  unit} $\eta$ may be defined through the unital element $1$ as 
$\eta (\xi) \rw \xi 1.$

2). Note that algebra (\ref{ancAlg}) is deformed  
 by  parameter $q$ as in the
  usual quantum algebra, but  additionally it is 
deformed also  by  central operators $\hat M^\pm $, prompting our 
definition of this deformed algebra as a novel
 {\it operator deformed} quantum algebra.
The deforming  operators being  central, 
they act   simply as multiplicative parameters,
which however   can take positive/negative  or
 zero values, generating different possible subalgebras as described
below. It should be noted  that for quadratic  algebra (\ref{ancAlg}) 
more sophisticated construction like quantum double, universal R-matrix
or even finite-dimensional representations might  be problematic and could
be     an interesting future problem to explore. However this does not 
create any problem  in applying it as a YB algebra 
for generating and classifying QIS, which is the main purpose of the present
investigation.
%SHESH
Generically this algebra allows infinite-dimensional representation, which 
%for nontrivial values of $\hat M^\pm $
 may be given through canonical
 operators $[u,p]=i $ as
 \be  s^3=u, \ \
s_q^+= e^{-i p}g(u),\  s_q^-=g(u) e^{i p}\ll{AArep}\ee
 where \be  g^2 (u)={1 \ov
2\sin^2 \al} \left ( \kappa + \sin \al (s-u) ( M^+ \sin \al
(u+s+1)-i { M^-
} \cos \al (u+s+1 ) ) \right ) \ll{g}\ee
with $\kappa$ and  spin variable  $s $ being
 arbitrary constants. We  drop $hat $ from the deforming central operators
 $\hat M^\pm $ which act only multiplicatively and  
 can 
also have zero values.
%central elements
%$ M^\pm=\pm \sqrt {\pm 1} ( c^+_1c^-_2 \pm
%c^-_1c^+_2 )

The integrable
 ancestor model, as obtained using   (\ref {AArep},
\ref {g})
in its  Lax operator (\ref {aL}), represents a generalized lattice
sine-Gordon (LSG) model. It reduces to the known LSG \c{lsg} in the absence of
 deforming operators (i.e. at  $\hat
M^+=1, \ \hat M^-=0 $), when  we also
 recover  from (\ref{ancAlg}) the  well known $sl_q(2)$ 
   and from (\ref{aL}) the  known Lax
operator of  \c{frt}.
In general (\ref{ancAlg}) can yield {\it nine} major subalgebras for the range of
values
$M^+ > 0, <0,=0 $ (up to  scaling), for each of the  sectors  $M^- =0, 0>, <0$ .
Among these  subalgebras    $M^+ > 0, \
M^- =0 $ yields the well known $sl_q(2)$, while $M^+ < 0, \
M^- =0 $  sector gives the corresponding noncompact case. All other cases
give a variety of subalgebras describing
 underlying YB algebras of important quantum
integrable systems, including new ones.

\subsection { Operator-deformed
 algebra at $q \to 1$-limit}
Consider now $q \to 1 $ limit of the above
 operator-deformed algebra, when  the related structures reduces to their
corresponding  rational limit:
Quantum  $ R^{trig}$-matrix  goes to
the rational one $  R^{rat}$. The 
 operators $s^b_q \to
s^b, b=1,2,3, $ \ $\hat c^\pm_a \to \hat c^{0,1}_a, a =1,2  $
(since they  are also
q-deformed!) and the
spectral parameter $\xi \to
\la $, while
 the ancestor Lax operator (\ref{aL})
reduces to its rational form
\be
L_{anc}^{rat}(\la) = \left( \begin{array}{c}
 \hat c_1^0 (\la + s^3)+ \hat c_1^1 \ \ \quad
  s^-   \\
    \quad
s^+    \quad \ \
\hat c_2^0 (\la -s^3)- \hat c_2^1
          \end{array}   \right). \ll{LK} \ee
 Finally algebra (\ref{ancAlg}), with     deforming operators
$ \hat M^\pm \to  \hat m^\pm $, reduces to
another novel operator-deformed (but  $ q \to 1$)  bi-algebra
\be [ s^+ , s^- ]
=  2\hat m^+ s^3 +\hat m^-,\ \ \ \
  ~ [s^3, s^\pm]  = \pm s^\pm , \ \ [\hat m^\pm, \cdot]=0
\ll{Csl2}\ee
with  central operators $\hat m^+=\hat c^0_1\hat c^0_2,
\ \hat  m^-=\hat c^1_1\hat c^0_2+\hat c^0_1\hat c^1_2$, as expressed through 
  mutually
commuting central operators $\hat c^a_i, \ a=0,1; i=1,2$.
Interestingly it  still exhibits
 non-cocommutative coproduct given by
\ $ \Delta(  s^+)=\hat c_{1}^{0 }\otimes s^+
                 +   s^+ \otimes \hat c_{2}^{0 } , \
 \Delta(  s^-)=\hat c_{2}^{0 }\otimes s^-
                 +   s^- \otimes \hat c_{1}^{0 } ,$ etc.
Note  that for nontrivial  $\hat  m^- $ this is a new operator-deformed
Hopf algebra   with {\it nine} distinct
subalgebras with
$m^+ > 0, <0,=0 , $ for each of the combinations from
  $m^- = 0, >0,<0 . $ The case
 $m^+ > 0, \
m^- =0 $ yields the standard   Lie algebra  $sl(2)$, while $m^+ < 0, \
m^- =0 $   yields  its noncompact variant $sl(1,1)$. The rest of the cases gives the
freedom of describing other integrable models.

\subsection{Application to quantum integrable systems }
Since only the ancestor Lax operator (\ref{aL})  contains  the deforming
operators but not the associated $R^{trig}$-matrix, we can construct
  Lax operators of all integrable systems
 belonging to the same trigonometric class together with
their YB algebras, by suitable
choices  of the deforming operators. Similarly at  $q\to 1 $ limit we would
get
the corresponding result for the models belonging to
 the rational class all sharing the same $R^{rat}$-matrix.
We identify three  types of integrable models
that we can  construct  following our scheme,
depending on different range of values of the deforming operators.

i) For   fixed  values of the deforming operators on the whole lattice
we get from (\ref{ancAlg}) its different  subalgebras which
interestingly represent the underlying YB algebras of
   original
integrable models. At the same time the ancestor
$L_{anc}^{trig}(\xi)$-operator
(\ref{aL}) and its rational limit $L _{anc}^{rat}$ (\ref{LK})
generate in a unified way
  the  representative Lax operators of known as well as new quantum
 integrable models as
 \\  $
R(xxz) \ : \ L_{anc}^{trig} \Longrightarrow  
%\left [ 
%\begin{array}{c}
i)  xxz - \mbox {spin chain}\ \ ,
ii)  \mbox {sine-Gordon (lattice +field)} \\ \ \  , iii)
 \mbox {derivative NLS (lattice +field )}  \ \ , iv)
 \mbox {q-bosonic}\ \ , v)
\mbox {Relativistic Toda chain} \\ \ \, vi)
\mbox {Liouville (lattice +field)}\ \ vii)
 \mbox {massive Thirring (bosonic)}  $
%\end{array} 
%  \right.$
%  \]
%\input{anc.eps}
\vskip .5 cm

 And at $ q \rightarrow 1 $ (rational limit)
%  \[
\\ $R(xxx) \ : \ L_{anc}^{rat} \Longrightarrow 
%\left [
% \begin{array}{c}
  i) xxx - \mbox {spin chain}\ \ ,ii)
  \mbox { NLS (lattice +field)}\\ \ \,iii)
   \mbox { simple lattice NLS }\ \, iv)
\mbox { Toda chain} $
% \end{array}  
% \right.$ 
% \]
%SHSEH$R^{trig}$

This systematic generation of
quantum integrable models from a single ancestor model also answers
to  some  intriguing questions raised above.
%asking why   different  models should share the same R-matrix  and
%what is the  relationship between these diverse models.
Our  finding shows clearly that all these
integrable  models are related deeply to each other as descendants
 of
the same ancestor and therefore they
 share   the  same inherited  R-matrix. 
 Moreover their underlying YB algebras
 are  obtained as different subalgebras of the same ancestor algebra.

ii) For variable values of the deforming operators
 on the lattice however
we obtain 
%inhomogeneous versions of the original integrable  models, yielding
  a new series of  quantum integrable inhomogeneous
models.
% In particular from
% $ L_{anc}^{trig}$  one can generate
%{\it  inhomogeneous  SG model} with variable mass,
%{\it Inhomogeneous quantum Ablowitz-Ladik model} etc, while from
% $ L_{anc}^{rat}$ we can construct
%{\it  inhomogeneous  NLS model}, {\it  inhomogeneous  quantum Toda chain} etc.

iii) Choosing   different  but  fixed  values of the
  deforming operators
at different sectors of the  lattice, we  can combine   different
interacting
integrable models and generate  a
novel class of  integrable   hybrid models.

% A striking example is the
%quantum integrable bosonic massive Thirring model, constructed by fusing
% two derivative NLS models. Other important examples
%are hybrid integrable models on the lattice,
%a range of integrable matter-radiation models including their
%  q-deformations and integrable field models with defect.

We present below  details of the YB algebras as
subalgebras of our ancestor algebra
 and its   major applications in  quantum integrable systems as
mentioned above.
\section{Generation of  integrable models and  underlying YB algebras}
Starting from the ancestor Lax operators (\ref{aL}) and  (\ref{LK}) and the
related operator deformed algebras (\ref{ancAlg}) and (\ref{Csl2})
we generate  quantum integrable models, known as well as new, together with
their underlying  YB algebras  in a
unified way.

\subsection{q-deformed algebra  and related integrable models}
We first focus on  models belonging to the trigonometric class with
$R^{trig} $.

1) At q-deformed but  operator-undeformed limit:
$ c^+_1=c^+_2=c^-_1=c^-_2=1$ giving $ M^+=1, M^-=0  $,
our general   algebra
  (\ref{ancAlg}) reduces  to the
  known quantum algebra  $sl_q(2) $ 
%(\ref{slq2})
 and the ancestor Lax operator (\ref{aL})
 to the  known $L(\la)$- operator of  \c{frt}. Consequently
 the spin-$\ha $ representation would  generate $xxz $-spin chain \c{fadrev}, while 
realization (\ref{g})  reproduces
$ g (u)={1 \ov 2 \sin \al}
  \left [ 1+ \cos \al (2 u+1)
 \right ]^\ha $ related to the known
 lattice SG model \c{lsg}.

 2) $M^+=-1, M^-=0  $ case reduces
  (\ref{ancAlg}) to the corresponding noncompact quantum algebra
 $sl_q(1,1) $ and reproduces the related integrable model, e.g. $q$-deformed
Buck-Sukumar(BS) model.

 3)  Taking the deforming operators  as
 $ \hat c^\mp_{1}=( \hat c^\mp_{2})^{-1}=p^{ \hat c},\ $
we get $p,q$-parameter-deformed  Hopf algebra
   $gl_{p,q}(2)$, which is represented by the
 same $sl_q(2)$ algebra, but is $ p^{ \hat c}$ -deformed
as a co-algebra. This YB algebra is  linked with
 the quantum integrable
 {\it Ablowitz-Ladik model}.
 
It is worth noting that, such an algebra is    believed to be
obtainable only through {\it twisting}
 transformation with an external parameter 
$p=e^{i\theta} $ \c{pq}.  Under such a transformation  the 
$R$-matrix is also deformed 
 to a twisted one: $R^{trig}(\la, \theta)$-matrix.    
However, contrary to this  belief  we obtain the same bi-algebra without 
changing the $R^{trig}(\la)$-matrix, since in our case the deforming
operators replaces the role of twisting parameter as mentioned above.
 
4) Choosing the values of the operators as  $ M^+= {\sin \al }
, \ M^-= i {\cos \al }$, compatible  with
$ c^+_1=c^+_2=1, \ c^-_1= -{iq }  , \ c^-_2=  {i \over  q}$
one gets  the $q$-bosonic  algebra
\be { [\psi_q,N] = \psi_q, \ \  [\psi_q^\dag,N] =- \psi_q^ \dag,\ \
 \  [ \psi_q, \psi_q^ \dag ] =
\cos (\al (2N+1)) }, \ll{qboson}\ee
by assuming $s_q^+=\psi_q, \ s_q^-=\psi_q^\dagger,\ s^3=N.
 $
This opens up an interesting possibility of constructing quantum integrable
models involving $ q$-boson \c{qbose06}.
In this case   (\ref{g})
  simplifies to $g^2(u)=
{[-2u]_q}$

A $q $-boson  can be mapped into
a standard boson $ \psi$   on the
lattice with  commutation relation
            $ [\psi , \psi^\dagger ] = {\hbar \over \Delta
}, $   as $ \psi_q =
     ~\psi (\frac{[2N]_q}{2N \cos \al})^{\ha}, \ N=
     \psi^\da\psi \ . $
 This can construct
   from the above q-bosonic model
an  exact lattice version of the {\it
quantum integrable derivative NLS}
(with   Lax operator obtained directly from  (\ref{aL})),
 which at the continuum limit $ \Delta \to 0$
yields the  quantum DNLS   model      \c{qdnls}
represented  by the equation
$ \  i \psi_t =  \psi_{xx}  -  i  (  \psi^\dagger  \psi  )  \psi_x . \ $
 This QFT model
 in the N-particle sector   is  equivalent
 to the interacting Bose gas with {\it derivative} $\delta$-function potential
\c{shirman}.

5) For another choice
%$ c^+_1=c^-_2=1, \ \ \ c^-_1=c^+_2=0$
$ M^+=M^-=1$  we derive a novel   {\it exponentially} deformed
 algebra
\be [ s_q^+, s_q^-]= {e^{2i\al s^3} \over 2 i \sin \al }, \ll{expalg}\ee
which turns out to be  the underlying YB  algebra of  the
    quantum integrable  exact lattice {\it Liouville model},
with its Lax operator   obtained from (\ref{aL}) with realization
(\ref{g}) reducing to
$  g(u)= {(1+e^{i \al(2 u+1)})^\ha \ov \sqrt {2} \sin \al}. $
At the  continuum limit it gives  the quantum  Liouville field
model
 given by the equation
$ \ u_{tt}- u_{xx} =  e^{\alpha u}  \ $

It is important to note that  the present values of  $ M^\pm$
  remain same  even with
$ \ c^-_1 \not =0$, which  would give
 the same  algebra and the  same realization,  but a different   Lax operator.
 This is indeed an intriguing
possibility of constructing different useful Lax operators for the same
model, in a systematic way. For example,
 a {\it second} Liouville Lax
operator  we  can construct here so easily  recovers
 that invented  by Faddeev
 through many innovative tricks \c{fadliu}.

 6 ). In a similar way the particular case $M^\pm=0$ can be achieved with
  different sets of choices:
$$
i) \ c_a^-=0, c_a^+=1,  \ a=1,2 \ ii) \ c_2^\pm=0, c_1^-=-c_1^+=1 ,iii) \  c_1^+=1,
\ \mbox {rest of c's }=0, $$
 all of which lead to the same   algebra
\be
[ s_q^+, s_q^-]= 0, \ [ s^3, s_q^\pm]= \pm s_q^\pm .\ll{nula} \ee
However, they may generate   different
  Lax operators  from (\ref{aL}),
 corresponding even to different  models, though
with  the same underlying algebra.  In particular, case i) leads to
  the {\it light-cone SG} model, while
 ii) and iii)   give two different  Lax operators  found in
\c{rtoda} and \c{hikami} for  the same  { relativistic
 Toda chain}.
Since  here we get
  $g(u)=$const. ,  interchanging $u \rw -ip, p\rw -i u,$
  in (\ref{g}) yields  simply
$ \ \
 s^3 =-ip, \ s_q^\pm=  \al e^{\mp u }, \ \ $
% \ll{dtl1} \ee
  which generates a quantum integrable      {\it discrete-time or relativistic
quantum Toda chain} \c{rtoda}.

 Remarkably,
 all  above algebras are  obtained
as  subalgebras  from  our ancestor algebra, but can not  be  obtained
 from $sl_q(2) $,
  without invoking singular limits.
Note also that all the descendant models  listed above
have the same trigonometric $R^{trig}$-matrix,
 inherited from the ancestor model and similar is true
  for   the rational class, as we will see below.  This unveils the
mystery why a wide range  of  models found to share the same
$R$-matrix, while their  $L$-operators are obtained
from the same  ancestor Lax operator
   (\ref{aL})
and their underlying YB algebras from   the same
ancestor algebra (\ref{ancAlg}) at various reductions.

\subsection{q-undeformed algebra  and related integrable models}

We focus now on the $ q\to 1$ limit of  algebra
(\ref{ancAlg}), which would still be
operator deformed through $\hat m^\pm $.
It is interesting to find that   the bosonic representation
 (\ref{g}) at this limit
  reduces to a generalized Holstein-Primakov transformation (HPT)
 \be
 s^3=s-N, \ \    s^+= g_0(N) \psi, \  \  s^-= \psi^\dag g_0(N)
, \ \ g_0^2(N)=\hat m^-+\hat m^+ (2s -N), \ \ N=\psi^\dag \psi.
\ll{ilnls} \ee
with the central deforming operators acting multiplicatively. 
This is   an exact realization of  (\ref{Csl2}),
associated with the Lax operator (\ref{LK}).
This rational  ancestor model,   representing  a quantum
  integrable {\it generalized  lattice NLS} model,
can generate  in a systematic way  all  other integrable models
   of the  rational class
 through different choices for the values of the deforming
operators. This constructs    at the same time the Lax operators of
the models as reductions of    (\ref{LK})  and their underlying YB
algebras as subalgebras of  (\ref{Csl2}).

1) When the deforming operators vanish with the choice
 $  m^+  =  1,m^-  =  0, $
  (\ref{Csl2}) gives clearly  the standard
   $su(2)$ and for the  spin $\ha$
representation one  recovers the  $xxx$ {\it  spin chain} \c{fadrev}.

2) A bosonic
 realization  of  the same algebra   simplifies (\ref{ilnls})
 to  the standard  HPT and (\ref{LK}), reproducing
the well known  {\it lattice   NLS} model
\c{lsg}.

3). On the
other hand for  $  m^+  =  -1,m^-  =  0, $
(\ref{Csl2}) reduces to  $su(1,1)$ and gives the
related integrable models, one of which is the integrable
BS type model constructed in sect. 7.

4). For the complementary  choice $  m^+  =  0, m^-  =  1, $
 (\ref{Csl2})  reduces to
    a non semi-simple algebra
\be
[ s^\pm, s^3]=\mp s^\pm, \ \   [ s^+,  s^-]= 1
\ll{bose} \ee
and (\ref {ilnls}) with  $g_0(N)=1$ gives a direct bosonic
realization
$s^ {+}=\psi, s^ {-}=\psi^\dag, s^ {3} = s -N .$
Using this reduction in (\ref{LK})  we  discover
 another quantum integrable {\it simple
lattice NLS} model
\c{kunrag}.

5)  A trivial choice $m^\pm  = 0$  reduces
(\ref{Csl2})   to  the same
  algebra (\ref{nula})
\be
[ s^+, s^-]= 0, \ [ s^3, s^\pm]= \pm s^\pm \ll{toda}\ee
and   therefore we can take the same
  realization $ \ \
 s^3 =-ip, \ s^\pm=  \al e^{\mp u } \ \ $
found  for the relativistic case, but   now  with  ancestor    Lax operator
(\ref{LK}) and  $R^{rat}$-matrix.
 This  gives quantum integrable
  {\it nonrelativistic
 Toda chain} \c{kulskly}.

Thus we have shown  how the ancestor model (\ref{LK})
with (\ref{ilnls}) can generate quantum integrable models, all sharing the same
rational $R $-matrix in a unified way
and their YB algebras
 like $ sl(2)$, (\ref{bose}), (\ref{toda} ) are obtained as  subalgebras
 from operator-deformed ancestor algebra (\ref {Csl2}).

It is important to  note that
   Lax operators like (\ref{aL}) and
({\ref{LK}) in their   bosonic realization  appeared  in  some
 earlier work  \c{taraskor} and they were shown to be the most general
possible form in their respective class.

Apart from the discrete models obtained above, one can
 construct integrable  QFT models from their exact  lattice versions
 at the  continuum limit. For such construction
we have to scale    the operators like
$ \Delta p_j, \Delta  c^\pm_a,
\Delta ^{1 \over 2}  \psi_j,$ with lattice spacing $\Delta$, which
at continuum limit
$\Delta \rw 0$ give $p_j \rw p(x), \psi_j \rw \psi(x)$ etc.
In this process  Lax operator ${\cal L}(x, \la)$ for the continuum model
is obtained
from its discrete counterpart as $L_j(\la) \rw I+ \De {\cal L}(x).$ The
associated $R$-matrix however remains the same since it does not contain
$\De$. Thus {\it integrable quantum field models} like sine-Gordon,
 Liouville, NLS or the derivative NLS can be  obtained from their
discrete variants constructed above.

\section{ Quantum integrable inhomogeneous  models}
%\setcounter{equation}{0}

%It is surprising that in spite of a volume of literature devoted to
In spite of intensive study of integrable inhomogeneous classical models, their
 quantum versions seem to remain 
unexplored, except perhaps the recent work \c{kun06}.
We can construct    a novel class of
quantum integrable   models as 
  inhomogeneous versions  of the original 
 integrable models listed  above. The idea of  
 construction is to take the values of the deforming
operators to be  site dependent functions, which would  replace
 $\hat M^\pm$ in  $g(u)$ (\ref{g})
by site dependent operators $ \hat M^\pm_j$. Consequently  
all $\hat c$'s  in  Lax operator
(\ref{aL})   would be  site-dependent
$\hat c_{j}$'s.

%In such models  values of  deforming operators  (including zero values)
%can vary arbitrarily   with site $j$ , which
However  the YB algebra in 
  inhomogeneous lattice models remains the same as in their original models 
with the same quantum
$R$-matrix. Physical interpretation of  such inhomogeneities may be as
impurities, varying external fields, incommensurate-Ness etc.
 Few examples of such models are

\noindent 1). {\it Inhomogeneous sine-Gordon model}:

 Taking the values of deforming  operators  different
at different lattice sites:
 $c^\pm_1= \Delta m_je^{i\al \th_j},
 \ \ c^\pm_2= \Delta m_je^{-i\al \th_j},$ which yields
 $   M_j^+=(\Delta m_j)^2,  \ M_j^-=0,$
 one can construct a novel {\it variable mass discrete
SG model} without spoiling its integrability.
 In the  field limit $\Delta \to 0 $
 it would yield a quantum integrable
  {\it inhomogeneous} sine-Gordon
 model with variable mass $m(x)$ in an  external gauge field $\th (x)$.

%In the simplest case the Hamiltonian of such model would be
%${\cal H}= \int dx \left ( m(x) (u_t)^2 + m^{-1}(x) (u_x)^2 + 8(m_0-m(x)
%\cos (2 \al u )) \right). $ Similar models
% may arise also in physical situations
%\c{msg}.

\noindent 2). {\it Inhomogeneous lattice NLS} model

Consider  site-dependent (in general also time-dependent)
 deforming operators  in (\ref{LK}) and in
 (\ref{ilnls}) as
\be c^0_1=c^0_2 \equiv g_j(t) , \
c^1_1=-c^1_2=f_j(t), \ \ \mbox{ giving}
\ m^+=g_j^2,\  m^-= 0, \ee
 with $f_j, g_j$   time dependent
 arbitrary discrete functions. This  gives  from (\ref{ilnls})
\ $ g_0^2(N)=g^2_j (2s_j -N_j  {\De }), $ which reduces
  (\ref{LK}) to the $L$-operator of the {\it quantum integrable
inhomogeneous }  exact lattice NLS. At   continuum and 
 high  spin limit
 $s_j \to {1 \over \Delta} g_j^{-1}$ limit 
% we proceed in analogy with the known lattice NLS \cite {lsg} 
 we get a  quantum
integrable  inhomogeneous NLS field model with two arbitrary
 functions $f(x,t),g(x,t) $
%, which can be chosen differently to produce
%wide range   of inhomogeneities
 \c{kun06}.
Note that
 unlike  conventional
 inhomogeneous classical models, our construction is
 iso-spectral, since inhomogeneity enters here solely through the deforming
operators in the Lax operator without affecting the spectral parameter and
hence the $R $-matrix.

\noindent 3).  {\it Quantum integrable inhomogeneous Toda chain}

We   choose inhomogeneity as  $  c^1_{1j},
c^0_{1j}$ with
$ c^1_{2}=c^0_{2}=0$, resulting   $m^\pm=0$.
%generate the   inhomogeneous extension of the quantum Toda
 This  gives from  (\ref{LK}) quantum
integrable inhomogeneous Toda chain
with the  Hamiltonian
\be H= \sum_j (p_j +{c^1_j \ov
 c^0_j})^2+{1 \ov c^0_jc^0_{j+1}} e^{u_j-u_{j+1}}
\ll{todah}\ee
%containing  inhomogeneity through arbitrary $c^a_{j} \ ,a=0,1$.

 In a similar way
 inhomogeneous versions of Liouville model, relativistic
Toda, Ablowitz-Ladik model etc. can be constructed.
%It should be remembered that the quantum $R $-matrix associated with the
%inhomogeneous models are the same as for their homogeneous counterpart.

 \section { Quantum integrable  hybrid models}
 A new class of
integrable models may be constructed combining different integrable models
which interact between themselves and preserve their integrability as a
 hybrid model.

The  idea of construction is to
take Lax operators $L_j^{(a)}$ of {
different} descendant models $a=1,2,3..$,  having the same $R$-matrix
(as listed above) and
insert them at {\it different} sites $j$
% constructing  the hybrid system
through
$ \  T(\la)= L^{(a)}_1L_2^{(b)}\cdots
L_N^{(c)}, \ $
 which satisfies  the QYBE
%, since each of the $L^{(a)}_j $ does that
  and hence  represents a quantum integrable
model. {Few interesting  examples are}:

\noindent 1). {\it Massive
Thirring model} : A spectacular example is a new bosonic version of quantum
integrable massive Thirring model \c{qdnls,aa}. It  can be constructed
 as a hybrid of two integrable 
DNLS models
% (by using  the antipode symmetry of the subalgebra)
 by  fusing two DNLS Lax operators :
     $L_{DNLS1}(\xi, \psi_1)\otimes L_{DNLS2}^{-1}({1 \ov \xi}, \psi_2)$
yielding the  Lax operator of the massive Thirring model
involving two bosonic fields $(\psi_1, \psi_2) $.

\noindent 2). {\it Integrable
matter radiation models}: Following the above idea
we may construct a series of important matter-radiation models and their
q-deformations
 as quantum integrable hybrid models \c{kunmr}. We furnish more details
in the next section.

\noindent 3). {\it Exotic Hybrid models}: We can construct in principle hybrid
integrable lattice models by  combining  Lax operators of any lattice models
belonging to the same class. For example
one can build  hybrids of

i) Toda chain and
discrete NLS model, \ \ ii ) Relativistic Toda chain  and the
 Ablowitz-Ladik model \ \ iii)
lattice sine-Gordon with Liouville  model
etc.

\noindent 4) {\it  Integrable quantum field models with defect}

It is challenging to extend the concept of hybrid models  to quantum
field models. A particular direction of this program is to
construct  integrable quantum field model with defect, e.g., to construct
 NLS,
 sine-Gordon (SG) or Liouville QFT model with defect.

  Corrigan et al \c{corrig} have applied similar
idea for constructing SG equation at $x<0, x>0 $ and  a defect
at $x=0$. Their approach is concentrated mostly on the soliton scattering
problem in analogy with the classical case.
 It is  however a challenge
to   construct and solve
 a genuine
 quantum integrable SG field model
with defect. We
  could make such  a breakthrough  very recently
by building a quantum integrable  exact lattice version of the
SG model with defect which
satisfies the QYBE and
 allows exact  solution  through algebraic Bethe ansatz \c{kunhab}.
%At the same time our discrete model  goes to  an  integrable
% SG field model with defect at the continuum   limit,
%with its  Lax operator
% yielding  all higher  conserved
%operators.
 We furnish some details in   sect. 8.

\section{Quantum integrable  multi-atom
 matter-radiation models and their q-deformations }

The strategy of construction is to build the monodromy matrix of the system
   by taking  a  combination of different  Lax operators:
  \
 $T(\la)=L^{s}(\la) \prod_j^{N_a}L_j^{S}(\la)$,
with  $L^{s}(\la)$ linked either to  the rational (\ref{LK}) or to the
trigonometric (\ref{aL}) ancestor model, while
   $N_a$-number of $L_j^{S}(\la)$ are  related to  the spin or  the q-spin
 model.
 By construction  $T(\la)$
 must satisfy the YB equation with $R^{rat}(\la)$ or  the $
R^{trig}(\la)$-matrix
 in the first  or  in
  the second case, respectively,  representing a
quantum integrable system in both the cases with
$\tau(\la)=tr\ T(\la) $, generating a commuting set of conserved operators.
In  our matter-radiation models, the
 subsystem described by  $L^{s}(\la)$ with the
bosonic or the q-bosonic  realization represents the single mode radiation
field, while     $L_j^{S}(\la)$ (in spin or  q-spin realization) represents
atoms, i.e.   the matter part of the model.
%We give more details on these models in the next section.

%Following  the   idea of  hybrid models outlined  above,  we build
%   the Hamiltonian
% of our unified matter-radiation (MR) system for the rational class as
%\bea
%H_{{ MR}}&=& H_d+ H_{ S s}+H_{S S}, \nonumber \\
%H_d&=&\omega_f s^3+ \sum_j^{N_a}  { \omega_a}_j S^z_j ,\ \ \
% H_{Ss }= \al \sum_j^{N_a}\left (
% s^+ S^-_j+ s^- S_j^+ +(c^0_1+c^0_2) s^3S^z_j \right )
%,\nonumber \\
% H_{SS}&=& \al \sum_{i < j} \left((c^0_1+c^0_2) S^z_iS^z_j+ c^0_1 S^-_iS^+_j+
%c^0_2 S^+_iS^-_j \right)
%\ll{imrh}\eea
%Here
% $H_{Ss }$
% describes  matter-radiation, while
% $H_{SS } $,
%matter-matter interactions.
% ${\bf S}_j, j=1,2,\ldots, N_a$ stand
% for an array of $N_a$ atoms, each with $2s+1$ levels
% and satisfy the $su(2)$ algebra. ${\bf s}$ on the other hand
%signifies  a radiation or a
%vibration mode and  satisfies more
% general algebra
%(\re {Csl2}). In
% (\re{imrh}) the  radiation frequency $\omega_f $ and the  atomic
% frequencies $\omega_{aj}, j=1,2, \ldots, N_a$
% are defined through  inhomogeneous
%parameters of the Lax
%operator as
%$ \ \omega_f=\sum_j w_j, \ \ w_j=\al (c^0_1-c^0_2) c_{j}
%, \ \ \  \omega_{aj}= \omega_f-w_j +\al (c^1_1+c^1_2)
%\ \ .$
%\ll{frec}\ee

Thus through different reductions of
a general    model    we can generate the following
 integrable multi-atom
MR models in a unified way \c{kunmr}.
% few examples of which are given below.

\noindent 1). {\it Integrable multi-atom  Buck-Sukumar (BS) model}:

%For the choice
% $ \ c_1^0=- c_2^0=1,  c_1^1=c_2^1 \equiv c,
%\ $ (\re{Csl2}) reduces to $su(1,1) $
%  \ll{cbs}\ee
%  and   (\re{imrh}) simplifies to
The model is given by $H=H_d+ H_{ S s}+H_{S S} $ with
\be
H_{BS}= \omega_f s^3+ \sum_j^{N_a} \left( { \omega_a}_j S^z_j+
 \al (s^+ S^-_j+ s^- S_j^+)\right)
+   \al \sum_{i < j}^{N_a} (  S^-_iS^+_j- S^+_iS^-_j),
 \ll{nbsh}\ee
 which with
a bosonic
realization of $su(1,1)$:  $ \
 s^+=\sqrt N b^\dag,
s^-= b \sqrt N , s^3=N+\ha \ $ and the spin-$s$ operator
${\vec S}=\ha \sum_k^{2s}{\vec \sigma}_k$,
 represents an  { integrable multi-atom
 BS model} with inter-atomic interactions and nondegenerate
atomic frequencies (AF).
At $N_a=1$, 
%when matter-matter interactions vanish and all  AF coincide,
  (\re{nbsh})
  recovers
   the known  BS   model  \c{bs}.
%by representing  ${\vec S}_1=\ha \sum_k^s{\vec \sigma}_k$.

\noindent 2). {\it  integrable multi-atom  Jayanes-Cummings (JC) model}:

%  Under reduction
%$\ c_1^0=\al, c_2^0=0, c_1^1 \equiv
%c, c_2^ 1=-\al^{-1}\  $
% (\re{Csl2}) reduces to $su(2) $
% and with its direct
With bosonic
realization
$s^-=b, s^+=b^\dag, s^3= b^\dag b , $
we get  a similar multi-atom
JC type model.
The known JC  model \c{jc} is recovered  at $N_a=1$,
  when interatomic couplings vanish and all AF  coincide.

\noindent 3).  {\it integrable trapped ion (TI) model}

%Taking  values of the deforming operators as $c_1^0=-1,
% c_1^1 \equiv  c, c_2^ 0= c_2^1=0  $, when
% (\re{Csl2}) goes to  (\re{toda})
with  realization
 $s^\pm=e^{\mp i x}, \ \ s^3=p+ x$,
% which  gives from the same
% (\re{imrh}) 
we get a new quantum  integrable trapped ion model
with  exponential nonlinearity, having the form
 (at $N_a=1 $)
\be
H_{TI}= (\omega_a-  \omega_f)S^z+{S^z}^2+
\al (e^{-ix} S^++e^{ix }S^-)+ H_{x p}, \ \
H_{x p}= \ha (p^2+ x^2)+ xp,
 \ll{tih}\ee
%with
% $H_{x p}= \ha (p^2+ x^2)+ xp,
$ \ {\vec  S} =\ha \sum_k{ \vec \sigma }_k, $
% which   is  a   new integrable multi-atom TI
%model with full exponential nonlinearity
% without  approximation.
% which is linked interestingly with the
% construction of the  well known Toda chain \c{kunprl}.

\subsection { Integrable q-deformed MR models}
 The strategy  here is   the same, only
  one has to start  now
from the  trigonometric  ancestor Lax operator (\ref{aL})
and  associated
$R^{trig}$-matrix.
The  Hamiltonian for our q-deformed MR models
 takes the form  (for  $N_a=1$)
  \be
H_{qMR}= H_d+  (s_q^+ S^-_q+ s^-_q S_q^+)\sin \al ,  \ \ \
H_d=-ic_0 \cos (\al X) +c \sin (\al X),
 \ll{qimrh}\ee
with $ \  X=(s_q^3-S_q^z+\omega), $ which represents  a new class of {  integrable
 MR models} with ${\bf S}_q \in su_q(2))$
 and   ${\bf s}_q$ generating
(\ref{ancAlg}). We  list below  a new series of  q-deformed
integrable MR models  which we obtain from the same  (\re{qimrh})
 for different realizations of
  ${\bf s}_q$.

\noindent 1). {\it Integrable  q-deformed   BS model}

This may be  constructed   from (\re{qimrh})  at $c_0=0 $,
 by realizing ${\bf s}_q$ through q-boson:\\
$s_q^+=\sqrt {[N]_q} b^\dag_q, \ s_q^-=b_q\sqrt{ [N]_q}, \ s_q^3=N+\ha  $,
and  quantum spin operator  ${\bf S}_q$
 through its  co-product : \\
$S^\pm_q=\sum_j^{s} q^{-\sum_{k<j}\sigma^z_k}
\sigma^\pm_jq^{\sum_{l>j}\sigma^z_l}, \ S^z =\sum_j^{s}\sigma^z_j$.
Note that at  $s=1$, we get an integrable
  version of an earlier  model \c{qoskul}.

\noindent 2). {\it Integrable q-deformed  JC model}

The model can be constructed similarly  from   the same
general model (\re{qimrh}) with   choice $c_0=i, c=1$ and
 q-bosonic realization $s_q^+= b^\dag_q, s_q^-=b_q, s^3=N. $

\noindent 3).  {\it Integrable q-deformed  TI
model}:

Under  reduction  $c_0=i, c=0$ and
 realizing through canonical operators:  $s_q^\pm=e^{\mp i x}, \
\ s^3=p, $ we can construct  a q-deformed
 TI model again from  (\re{qimrh}).

By taking higher $N_a$ values multi-atom integrable variants
 of  the above
q-deformed matter-radiation models can be constructed.
% related to the  relativistic Toda chain \c{kunprl}.

%shesh11-8

\section{Quantum integrable sine-Gordon field model with defect}

At  the discrete level, as discussed above, it is easy to build
 hybrid models inserting  Lax operators of different integrable systems at
different lattice sites.
However when we try to 
take the  continuum limit
 for constructing  the corresponding field model, the above construction
breaks down due to the lack
of any overlapping smooth region.
Therefore one needs certain mechanism similar to the classical formulation
of the same problem \c{corrig}.

We report here about some  breakthrough  in solving this problem,
with  details to be presented
separately \c{kunhab}.
 We construct  our  quantum integrable sine-Gordon model with defect
 first as an exact lattice version  with  its monodromy matrix  defined as
\be
T^{N}_{-N}(\xi)=\left(L^{+}_N(\xi, u^+_N) \cdots L^+_1(\xi, u^+_1)\right)
 F^d_0(\xi,u^+_0,  u^-_0)
\left(L^-_{-1}(\xi, u^-_{-1})\cdots L^-_{-N}(\xi, u^-_{-N})\right)
 \ll{dsgT} \ee
  where $L^\pm_j(\xi, u^\pm), \ j=\pm 1, \ldots \pm N $
% and $L^-_i(\xi, u_i), \ i=-1, \ldots -N $ 
is the quantum
 Lax operator  of  exact
lattice SG model \c{lsg} for the  field $ u^\pm$.
 At the defect point $j=0$ we define an  overlapping quantum Lax
operator
\be
F^d_0(\xi,u^+_0, u^-_0) = \left( \begin{array}{c}
  \xi   e_-^{-1} (P^+)^{-1} \qquad \ \
a e_+^{-1} P^- \\
    \quad
-ae_+  (P^+)^{-1}  \qquad \ \  \xi   e_- P^+
          \end{array}   \right),  \ll{dsgF} \ee
where $ \
  e_\pm= e^{\pm i {\al \over 4} (u^-_0\pm  u^+_0)} ,\ P^\pm e^{ {i \over 2} 
(p^-_0\pm p^+_0)}  \    $, with $
[u_k,p_j ]= i\delta _{kj}$.
Remarkably  (\re{dsgF}) is linked with
the  ancestor Lax operator (\ref{aL}) under reduction $c^a_-=0, a =1,2, $ and
therefore must satisfy the QYBE (\ref{qybe}) with
 $ R^{trig}$ matrix, which is also the case
with the sine-Gordon Lax operators $L^\pm_j$ appearing in (\re{dsgT}).
This proves  the quantum integrability of the system with
$\tau(\xi)=tr( T^{N}_{-N}(\xi)), $ generating the mutually commuting set of
  conserved operators.

It is crucial that this discrete integrable system  has a
 smooth continuum limit:  $\Delta \to 0 $,
with the fields $u^\pm_j \to u^\pm(x), p^\pm_j \to \Delta p^\pm(x)  $,
 having
$[u^\pm(x),p^\pm(y) ]= i\delta(x-y) $ and
$\sigma^1L^\pm_j \to (1 +\Delta U^{SG\pm}(x)), $ for $ x \in (0^+, +\infty)$
and $  x \in ( -\infty, 0^-) $, respectively, with $U^{SG\pm}(x) $ giving
the Lax operator for the sine-Gordon field model with fields $ u^\pm (t,x)$.
  At the defect point we get the transition:
  $F^d_0 \to (1 +\Delta {\cal L}(x))F_0  $, at
 $ x \in [0^-, 0^+]$, where
${\cal L}(x)= (F_{0,x}+F^{p})(x) F_0^{-1}(x)$, where $F^{p} $
 is obtained from $F^{d} $ by replacing $P^{\pm} \to {i \over 2} (p^-(x)
\pm p^+(x))$
%Backl\"und  gauge
and  $F_0$ gives  the known
Backl\"und operator  for the sine-Gordon model \c{hab}.
% is  obtained from our quantum integrable Lax operator $F^d_0$, when
% its factors $  P^\pm $ are dropped out at the continuum limit 
%due to scaling  of $ p^\pm_0 \to \Delta p^\pm$.

 Starting from the Lax operators at different regions we 
 can find all conserved quantities including  Hamiltonian, momentum etc.
  defined on the whole axis, 
(including the defect point) by 
 using the standard  Riccati equation technique. We can also find the
corresponding eigenvalues  exactly
 by applying  the algebraic
 Bethe ansatz method \c{fadsg} to the
 lattice regularised model (\re{dsgT})  we have
started with.

\section{Concluding remarks}
We have presented here a unified construction of quantum integrable models
from a single ancestor model based on a general Yang-Baxter Hopf algebra,
 representing  a new operator deformed quantum algebra.
Though some of our results were published earlier, we present our whole scheme
in a systematic way focusing on
 a new  concept  of  operator-deformation of an algebra, which is our
main concern. We also report on
some completely new result on the quantum integrable
sine-Gordon model with defect and other hybrid models like integrable
q-deformation of a few matter-radiation models.

We emphasize that  all integrable
 models listed here allow
  exact  Bethe ansatz    solution. Moreover
, similar to their unified construction
 one  can  get   their solution  also  in a unified and
   almost   model-independent way. For this one has to  start
 from the general solution of  the
 ancestor model following the algebraic
Bethe ansatz and   subsequently obtain the results for all descendant
 models with the same $R$-matrix
as various reductions of the general result.
\section {acknowledgement}
I express my sincere  thanks to the organizers of the NLPTE4 conference and
 the
 AvH foundation (Germany)  for  logistic and   financial support.

 \end{document}